\documentclass[lettersize,journal]{IEEEtran}

\usepackage{array}
\usepackage{booktabs}
\usepackage{comment}
\usepackage{makecell}
\usepackage{pifont}
\usepackage[caption=false,font=normalsize,labelfont=sf,textfont=sf]{subfig}
\usepackage[most]{tcolorbox}
\usepackage{todonotes}
\usepackage{xcolor}
\usepackage{hyperref}

\definecolor{oceanboatblue}{rgb}{0.0, 0.47, 0.75}

\newcommand{\totDetectors}{six}
\newcommand{\patchesPetke}{819}

\newcommand{\patchesSilva}{170}

\newcommand{\bugsPetke}{92}
\newcommand{\bugsSilva}{50}

\newcommand{\repo}{\cite{repo}}

\newcommand{\toolAcronym}{POD}
\newcommand{\dataOne}{\textit{classical}}
\newcommand{\dataTwo}{\textit{repairllama}}

\begin{document}

\title{Unveiling Practical Shortcomings of Patch Overfitting Detection Techniques}

\author{David~Williams,
Ioakim~Avraam,
Aldeida~Aleti,
Matias~Martinez,
Justyna~Petke,
and~Federica~Sarro%
\thanks{David Williams, Ioakim Avraam, Justyna Petke, and Federica Sarro are with the Department of Computer Science,
University College London, London, UK
(e-mail: \{david.williams.22,j.petke,f.sarro\}@ucl.ac.uk).}
\thanks{Aldeida Aleti is with the Department of Software Systems \& Cybersecurity,
Monash University, Melbourne, Australia
(e-mail: aldeida.aleti@monash.edu).}
\thanks{M. Martinez is with the Universitat Polit\`ecnica de Catalunya, Barcelona, Spain
(e-mail: matias.martinez@upc.edu).}
\thanks{This work has been submitted to the IEEE for possible publication. Copyright may be transferred without notice, after which this version may no longer be accessible.}
}

\maketitle

\begin{abstract}
Automated Program Repair (APR) can reduce the time developers spend debugging, allowing them to focus on other critical aspects of software development.  Automatically generated bug patches are typically validated through software testing. However, this method can lead to patch overfitting, i.e., generating patches that pass the given tests but are still incorrect. 

Patch correctness assessment (also known as overfitting detection) techniques have been proposed to identify patches that overfit. However, previous work often assessed the effectiveness of these techniques in isolation
and on datasets that do not reflect the distribution of correct-to-overfitting patches that would be generated by APR tools in typical use; thus, we still do not know their effectiveness when used in practice.

This work presents the first comprehensive benchmarking study of several patch overfitting detection (POD) methods in a practical scenario. 
To this end, we curate datasets that reflect realistic assumptions (i.e., patches produced by tools run under the same experimental conditions). Next, we use these data to benchmark \totDetectors{} state-of-the-art POD approaches -- spanning static analysis, dynamic testing, and learning-based approaches -- against two baselines based on random sampling (one from previous work and one proposed herein).

Our results are striking: 
Simple random selection outperforms all state-of-the-art POD tools for 71\% to 96\% of the cases, depending on the POD tool.
This suggests two main takeaways: (1) current POD approaches offer limited practical benefit, highlighting the need for novel techniques; (2) any POD approach must be benchmarked on realistic data and against random sampling to prove its practical effectiveness.
To this end, we encourage the APR community to continue improving POD techniques and to adopt our proposed methodology for practical benchmarking while doing so; we make our data and code available to facilitate such adoption.
\end{abstract}

\begin{IEEEkeywords}
    Automated Program Repair, Patch Correctness Assessment, Patch Overfitting Problem
\end{IEEEkeywords}

\section{Introduction}
Software bugs are an inescapable cost of software development, and debugging them consumes a significant amount of developer effort. 
Automated Program Repair (APR) has emerged as a promising technology to alleviate this burden by automatically generating bug \textit{patches}~\cite{asr_bibliography}. 
In practice, major companies~\cite{meta_apr_1,meta_apr_2,bloomberg_apr_1,bloomberg_apr_2} have explored integrating APR tools into their development workflows with encouraging results. 
However, a fundamental barrier impedes the wider adoption of APR: patch reliability. 
An APR-generated patch may pass all the available tests (making it plausible), but still fail to fix the underlying bug. 
Such patches are said to be \textit{overfitting} --- they ``fit'' the test cases but do not generalise to untested scenarios.

To bridge the gap between APR’s potential and its reality, the software engineering research community has proposed a range of \textit{Patch Overfitting Detection (POD)} approaches, also known as \textit{Patch Correctness Assessment}.
These techniques aim to automatically predict whether a given APR-generated patch is correct or an overfit. 
Various approaches have been explored~\cite{fei-2025-pca}, from dynamic analysis that executes patched programs on additional tests, to static analysis of code and semantics, as well as learning-based classifiers trained on past fixes. 
These tools promise to serve as an ``automated filter'', alerting developers to likely buggy patches so that only high-quality fixes are considered. 
However, in their survey, Fei~et~al.~\cite{fei-2025-pca} found that evaluations of POD tools to date have been isolated, often using differing benchmarks and custom criteria, making it difficult to gauge how existing proposals compare to each other or how they would perform in practice. 
Further, they highlighted that existing benchmarks are constructed by running multiple APR tools under inconsistent environments and budgets, reusing prior datasets, introducing human-written patches, or merging data from any combination of these sources.
While these evaluations are valid for assessing the raw classification performance of POD tools, they fall short of demonstrating whether the tools would provide practical value in contexts where end users would deploy them.
To address this gap in the literature, Petke et al.~\cite{petke_patch_overfitting_problem} proposed a novel methodology to evaluate practical POD tool benefits and demonstrated that when picking at random from a set of plausible patches, developers should expect to review only 2 patches for a representative bug to be confident they have encountered a correct one (if it exists).

In this work, we adopt Petke et al.'s methodology to carry out the first empirical study assessing the performance of \totDetectors{} representative state-of-the-art POD techniques spanning the major families of approaches (static, dynamic, and learning-based). 
For this, we curate two benchmark datasets of patches following stringent criteria.
Within each benchmark, each patch was generated under the same environmental setup, with each APR tool given the same time budget.
Using this approach, we can observe the performance of POD tools on a realistic distribution of correct-to-overfitting patches, as they are naturally generated by current APR tooling.

A key focus of our evaluation is the developer’s perspective on the patch inspection effort. 
We measure how many candidate patches a developer would need to inspect to find a correct one using each POD technique. 
To provide a baseline performance threshold for this measure, we compare POD tool performance with two theoretical baselines.
The first is the RS method advocated by Petke~et~al.~\cite{petke_patch_overfitting_problem}, which simulates a developer randomly picking patches to examine; this yields a success rate that any sophisticated method should beat.
Additionally, we introduce a novel second baseline, the Weighted Probability Classifier (WPC), a patch-agnostic method that randomly classifies patches while assigning a higher probability to the ``overfitting" class, reflecting the empirical observation that overfitting patches are more prevalent than correct ones. 
This distribution-aware strategy represents the best that a naive detector can achieve by leveraging only the base rate of correct patches. 
By comparing state-of-the-art POD tools against these baselines, we assess how much POD tools reduce the burden of identifying a correct patch.

Our results show that current detectors embody complementary virtues: learning-based approaches are fast and excel at discarding overfits, while dynamic analysers successfully classify correct fixes, but are slow and imprecise. 
For most bugs, however, our random baselines match or outperform them all.
For practitioners, our findings urge caution: Without further empirical evidence, one cannot assume that state-of-the-art POD tools will automatically reduce the manual debugging effort when incorporated into an APR workflow.
For the research community, the implications are equally clear: POD techniques must be evaluated under realistic conditions and against baseline benchmarks.
Our findings highlight that improving precision and recall simultaneously is the core of this challenge. 
We invite the APR community to build upon our work to ensure that novel POD techniques truly push the state-of-the-art in a way that matters for real-world use.
We provide a public, open-source replication package \repo{} to facilitate the adoption of realistic benchmarking for POD techniques.\looseness=-1

In summary, our study provides:
\begin{itemize}
\item The first empirical evaluation of POD approaches conducted on benchmark data reflecting realistic correct-to-overfitting patch distributions curated by mimicking real-world usage of APR tools.
\item A comparison of six state-of-the-art POD tools with two intuitive baselines --- Random Selection and Weighted Probability Classification (the latter being a contribution in itself) --- to contextualise the performance of POD techniques.
\item A robust and reusable evaluation methodology that establishes a fair basis for assessing future POD techniques.
\end{itemize}

\section{Background and Related Work}
\label{sec:relwork}
Traditional APR techniques~\cite{aprsurvey} rely on \textit{generate-and-validate} and \textit{semantics-driven} approaches to automatically validate generated patches without requiring human involvement.
The former explores a search space of candidate patches, validated via test executions, whereas the latter uses constraint solving to synthesise candidate repairs. 
More recent approaches, such as RepairLLaMA~\cite{repairllama2023}, leverage Large Language Models (LLMs) to generate patches using in-context learning. These methods offer significant speed improvements, as they can synthesise fixes in seconds.
However, these techniques still require robust validation to combat hallucinations and depend on pretraining/fine-tuning on large code corpora to achieve high-quality repairs, which introduces substantial upfront computational costs.

Given a buggy program and a specification of the correct behaviour, i.e., an \textit{oracle}, APR tools generate candidate \textit{plausible} patches that satisfy the oracle. 
The most common oracle used by APR tools is the target software's test suite.
Due to incompleteness of testing, APR-generated patches (i.e., plausible ones) are not guaranteed to be correct - such patches are said to \textit{overfit} to the given test suite~\cite{smith2015cure}. 
This phenomenon is known as the \textit{patch overfitting problem} in APR~\cite{le2018overfitting,long2016analysis,martinez2017automatic}. Among others, Ye et al. conducted a comprehensive study~\cite{ye2021quixbugsoverfittinganalysis} of patches generated by ten distinct APR tools, encompassing the three major tool categories (search-based, constraint-based, and learning-based). The results confirmed previous findings that the majority of patches generated were overfitting, regardless of the APR methodology used~\cite{liu-et-al-dataset, qi2015analysis}.\looseness=-1

To address the patch overfitting problem, overfitting detection techniques have been proposed~\cite{fei-2025-pca}:

1) \textit{Static Methods}
 perform overfitting detection by analysing the syntactic and semantic content of patches. They rely on the detection of anti-patterns in a patch to flag it as overfitting.
Such anti-patterns involve substantial structural divergence from the original code, violations of naming conventions, or other semantic violations that are hypothesised to correlate with incorrect behaviour. To detect anti-patterns, static approaches extract relevant features or representations from both the original and patched code, and leverage heuristics such as code similarity metrics or code embeddings to determine whether a patch exhibits overfitting characteristics.

2) \textit{Dynamic Methods}
 assess patch correctness by executing the program under patch in various scenarios and analysing its runtime behaviour. These methods typically extend the patch validation phase of APR systems: after a patch passes the given test suite, dynamic methods run additional tests or analyses on the patched program. 
Common dynamic approaches include automated test case generation, runtime monitoring for crashes or invariant violations, and execution trace comparison. 
The core assumption of dynamic methods is that a correct patch will not significantly alter the program’s behaviour beyond fixing the bug. Accordingly, dynamic techniques aim to uncover faults by augmenting the initial oracle, attempting to reveal counterexamples that expose overfitting patches. 
Unlike static methods, dynamic ones can directly detect logical errors in a patch through execution, at the cost of significantly higher computational overhead.

3) \textit{Learning-based Methods} train machine learning models, often using data from past patches, to predict patch correctness. 
The general workflow involves collecting a large corpus of patches from APR tools, labelling them, training a model to distinguish between correct and overfitting patches, and then applying that model to new patches. Many such techniques use neural networks to automatically learn a representation of the code change. Others use models directly in inference by prompting them with patch information to classify the patch.
The benefit of learning-based methods is that they can catch a wide array of issues that do not necessarily convey trivial static signatures. However, they heavily depend on the quality and representativeness of training data. Without diverse training data, they may not generalise to patches from new domains.

We refer the reader to Fei~et~al.'s survey~\cite{fei-2025-pca} for examples of the three aforementioned approaches.
A key takeaway from their survey was that determining the best overfitting detector is hard due to a ``\textit{lack of consistency in the datasets used and the evaluation focus of each method}'' (see also Section~\ref{sec:motivation}). 

\section{Motivation and Approach}
\label{sec:motivation}
Our key aim is to provide a comprehensive evaluation of state-of-the-art POD approaches within a \textit{realistic} scenario where a software engineer runs an APR tool (or multiple tools in parallel) up to a reasonable time limit; then each patch generated per bug is evaluated by an approach that detects overfitting patches. This will enable us to provide guidance on POD tools \textit{in practice}, filling the gap in existing work.

\subsection{Datasets for Evaluation of Overfitting Detection Methods}
\label{sec:current-overfitting-detection-evaluation-techniques}
The datasets used to date for evaluating patch correctness assessment approaches are typically assembled through one or both of the following methodologies: executing multiple APR tools on benchmarks such as Defects4J~\cite{defects4j}, regardless of whether they are given the same budget (e.g., set to stop at the first patch or not), then merging their results together, or reusing existing datasets, e.g., by combining sets of patches from replication packages associated with previous studies (which is the most commonly used approach so far~\cite{fei-2025-pca}). Some studies even include human-written patches in their datasets to enhance the representation of correct patches, thus balancing the dataset~\cite{DL4Patch}. Alternatively, others opted for the synthetic minority over-sampling technique (SMOTE)~\cite{smote}, to ensure an equal number of correct and overfitting patches when testing. 

When datasets are constructed from patches from different sources and experimental environments (such as those by Le~et~al.~\cite{8812054}, Durieux~et~al.~\cite{durieux2019empirical}, and Tian~et~al.~\cite{tian2022answer}), they may not reflect realistic patch distributions naturally produced by APR tools in practice.
While these datasets are valid for assessing the raw classification performance of POD tools, they fall short of demonstrating whether these tools would be viable in practical settings where developers would adopt them.
Consequently, to objectively evaluate their potential value in practice, POD tools should be evaluated on benchmark datasets where individual or multiple APR tools are allowed to execute within the same time limit, in a consistent computational environment~\cite{petke_patch_overfitting_problem}.
We thus curate two such datasets (Section~\ref{sec:dataset}) and make them available~\repo.

\subsection{Baselines for Overfitting Detection Methods}
\label{subsec:baselines}
Petke et al.~\cite{petke_patch_overfitting_problem} were the first to point out the lack of realistic evaluation of POD tools and performed a preliminary analysis on the occurrence of overfitting by executing 10 different APR tools on Defects4J \cite{defects4j} under the same experimental conditions. 
At the three-hour mark of their experiment, 155 bugs had at least one patch generated by one of the ten APR tools.
When analysing the dataset of patches generated at this point and removing duplicates, the \textit{median number of unique} patches generated per tool, per bug was 2, while the \textit{mean} was 6.71.
These findings suggest that if POD tools fail to demonstrate a significant predictive success, manual analysis by developers might not only be more accurate but also more time-efficient, given the limited number of patches to review.
Moreover, this study confirms the need for benchmarking POD techniques against realistic data and baselines. We thus consider two baselines outlined as follows.

\subsubsection{Random Selection (RS) Baseline}\label{sec:rs-baseline} 
This baseline, recommended by Petke~et~al.~\cite{petke_patch_overfitting_problem}, calculates the probability of identifying at least one correct patch through random sampling without replacement from a set of generated patches, given by:
\begin{equation*}
\label{eq:rs_baseline}
\resizebox{0.95\hsize}{!}{%
$
\begin{split}
Pr(X \geq 1) &=  1 - Pr(X = 0)  
= 1 - \frac{\binom{N-K}{n}}{\binom{N}{n}}\\
Pr(X \geq 1) & : \text{Probability of selecting at least one correct patch} \\
    N & : \text{Total number of patches} \\
    K & : \text{Total number of correct patches} \\
    n & : \text{Number of consecutive draws (patch selections)}\\
\end{split}
$}
\end{equation*}

\subsubsection{Weighted Probability Classification (WPC) Baseline}\label{sec:wbc}
We note that the RS baseline does not account for the intrinsic imbalance between overfitting and correct patches~\cite{liu-et-al-dataset}. 
Thus, we propose a naive method: Weighted Probability Classification. It explicitly accounts for this inherent class imbalance.\looseness=-1

Let $\text{B}(p)$ denote a stochastic classifier that predicts \emph{overfitting} with probability \(p\) and \emph{correct} with probability \(1-p\), independently for each patch, for $p\in [0.50,1.00]$: 
$Pr(\text{predict overfitting}) \;=\; p, \ \
Pr(\text{predict correct}) \;=\; 1-p.$  
Here, $p=0.5$ is equivalent to an uninformed coin flip, while $p=1.0$ corresponds to an ``always overfitting” guesser, which trivially achieves perfect negative recall (1.0) but offers no information on the correct-patch class. Sweeping $p$ across the full range $[0.5,1.0]$ produces an \emph{envelope} of expected performance.
Assume a dataset contains \( N \) patches, of which \( K \) are correct and \( N - K \) are overfitting. The proportion of correct patches is given by \(\pi = K/N\). The \textit{expected} confusion‐matrix rates for \(\text{B}(p)\) are:
\[
\resizebox{0.95\hsize}{!}{%
\begin{minipage}{\hsize}
\begin{align*}
\text{TP}(p) &= (1-p)\,\pi, &
\text{FP}(p) &= (1-p)\,(1-\pi),\\[2pt]
\text{FN}(p) &= p\,\pi, &
\text{TN}(p) &= p\,(1-\pi).
\end{align*}
\end{minipage}
}
\]

Substituting these expressions into any metric formula gives the \emph{expected} baseline value as a smooth function of~\(p\). 
One could in principle estimate~\(p\) from historic APR outputs, but such an estimate is tool‐, project‐, and time‐dependent and therefore not guaranteed to generalise. To remain method‐agnostic we evaluate the \emph{entire} range \(p\in[0.5,1]\) for insight, but we do not require that any candidate detector outperforms \emph{every} member of that baseline family. Instead, we classify each metric by whether a tool’s outcome lies above the baseline envelope for all $p$ (fully outperformed), some contiguous subrange of $p$ (partially outperformed), or no part of the range (not outperformed). 
When used together, the RS baseline informs developers of how effective a tool is in filtering out overfitting patches, and the WPC baseline provides a classification performance that any sophisticated POD method should surpass.\looseness=-1

\section{Methodology}

We evaluate the performance of state-of-the-art POD tools on data that mimics a practical use scenario, and ask:\looseness=-1

\textbf{RQ1: How effective are POD techniques at classifying patches produced by APR tools in practice?}
To answer RQ1, we selected \totDetectors{}  POD tools (two per methodological category -- static, dynamic, and learning-based) as representatives of the current state-of-the-art. We evaluated their performance using several metrics (Section~\ref{sec:metrics}) computed over all patches,  per bug, and per APR tool, on datasets we curated to meet practical scenario criteria (Sections~\ref{sec:current-overfitting-detection-evaluation-techniques}~\&~\ref{sec:dataset}).  This granularity allows for comparison of technique classes, revealing possible bug- or tool-specific insights.
   
\textbf{RQ2: How efficient are POD techniques on realistic APR data?}
To answer RQ2, we analyse the execution times of the six POD tools, aiming to assess effectiveness versus costs and identify cases where a given POD tool (or methodological category) might be more advantageous than others.

\textbf{RQ3: To what extent do POD tools provide practical advantages over naive baselines in terms of patch review burden and classification quality?}
To answer RQ3, we assess the performance of the detectors against the RS and WPC baselines (Section~\ref{subsec:baselines}) both across complete datasets and at the bug level.  
First, we quantify developer effort by counting, for every bug, the number of candidate patches a tool leaves for inspection before the first fix appears. 
These counts are contrasted with the theoretical workload of two RS baselines, RS-85 and RS-95, which model the number of random inspections required to achieve 85\% and 95\% confidence of encountering a correct patch. 
Second, we assess classification quality by computing the metrics outlined in Section~\ref{sec:metrics}, together with 95\% bootstrap confidence intervals, and overlaying each estimate on the performance envelope traced by the WPC baseline as the prior probability of overfitting is swept from 0.5 to 1.0.\looseness=-1

\subsection{Evaluation Metrics}
\label{sec:metrics}

There are four possible outcomes in patch classification: i) True Positive (TP): A correct patch is classified as correct; ii) True Negative (TN): An overfitting patch is classified as overfitting; iii) False Positive (FP): An overfitting patch is classified as correct; iv) False Negative (FN): A correct patch is classified as overfitting.

We report on the following metrics: $\text{Accuracy} = \frac{TP + TN}{TP + TN + FP + FN}$, $\text{Precision} = \frac{TP}{TP + FP}$, $\text{Positive Recall} = \frac{TP}{TP + FN}$, and $\text{Negative Recall} = \frac{TN}{TN + FP}$. 
Since Accuracy can be misleading under class imbalance, we also consider balanced accuracy, which weights each class equally.
We also consider the 
$\text{F1-score} = 2 \cdot \frac{\text{Precision} \cdot \text{Positive Recall}}{\text{Precision} + \text{Positive Recall}}$, 
 which reflects how well a tool identifies correct patches (Positive Recall), while at the same time penalising it for misclassifying overfitting patches as correct (Precision). 
This is particularly important when both FP (mislabeled overfitting patches) and FN (missed correct patches) carry meaningful consequences.

Since realistic patch datasets contain a higher proportion of the negative class (overfitting patches), Precision, Positive Recall and F1-score are useful as they focus on the classifier’s performance on the minority positive class, ensuring that the evaluation emphasises the accurate identification of correct patches despite their rarity. However, due to this class imbalance, each of the above metrics can be misleading when viewed individually, especially in the case of trivial majority classifiers or similar naive approaches~\cite{moussa}. 
For example, in the case of Accuracy and Negative Recall, the WPC baseline predicts the overfitting class more frequently than the correct one, and thus would achieve high scores for these two metrics.
This would result in a tool that discards most patches, leaving developers with a few patches left to observe that are likely overfitting.
To address this limitation, we use the Matthews Correlation Coefficient (MCC)~\cite{mcc} metric, which has been shown to be more robust to class imbalance \cite{mcc_binary,mcc_defect,moussa}:
$\text{MCC} = \frac{TP \cdot TN - FP \cdot FN}{\sqrt{(TP + FP)(TP + FN)(TN + FP)(TN + FN)}}$.
MCC ranges in $[-1, 1]$, where 1 and -1 signify perfect classification and misclassification respectively, and 0 corresponds to the performance of random classification.
Using MCC is highly informative as it yields a high score only if a patch correctness assessment tool can correctly predict both the minority positive class of correct patches and the majority negative class of overfitting patches.
Since MCC is undefined when a confusion‑matrix row or column is zero, we adopt a small‐value ``smoothing” as proposed by Chicco and Jurman \cite{mcc_binary}. In practice, we replace each zero cell by a tiny constant \(\epsilon = 1\mathrm{e}{-12}\), compute MCC, and recover a well‐defined score that tends to 0 in the cases of trivial classification.\looseness=-1

\subsection{Datasets} \label{sec:dataset}
To assess the extent to which a POD approach would provide a practical benefit to adopting developers, we need a dataset that reflects the scenario described in Section~\ref{sec:current-overfitting-detection-evaluation-techniques}.
Thus, we identify the following criteria that our datasets must meet.
First, the APR tools used to generate patches must be run using the same computational environment and time budgets, so as to reflect the natural balance between correct and overfitting patches generated by those tools, which end users (developers) would expect to see in practice. 
Second, patches must be labelled as correct or overfitting, ideally by human experts.
We further constrain our dataset search to Java programs in the Defects4J dataset, as it is the most popular benchmark for POD technique evaluations~\cite{fei-2025-pca}, and allows us to control for the nature of the bugs and programming-language context when making cross-dataset comparisons.
In the literature, we found only two datasets meeting our criteria: one by Petke et al.~\cite{petke_patch_overfitting_problem} containing patches generated by 10 traditional APR tools, and another by Silva~et~al.~\cite{repairllama2023} containing patches generated by RepairLLaMA, an LLM-based approach that demonstrated superior performance over zero-shot ChatGPT-based program repair. 
Table~\ref{tab:datasets} shows the distributions of correct-to-overfitting patches and other stats.

\begin{table}[t]
    \centering
    \caption{Datasets used in our experiments. The APR tools in each respective dataset are given the same budget and run in the same environments to simulate practical patch distributions.}
    \resizebox{0.95\columnwidth}{!}{
    \begin{tabular}{lrrrl}
      \toprule
      \textbf{Name} & \textbf{\# Patches (Corr. : Overfit.)} & \textbf{\# Bugs} & \textbf{\# APR Tools} & \textbf{Notes} \\
      \midrule
      \textsc{classical}~\cite{petke_patch_overfitting_problem} & 819 (129 : 690) & 92 & 10 & Defects4J; 8h budget \\
      \textsc{repairllama}~\cite{repairllama2023} & 170 (63 : 107) & 50 & 1 & Defects4J; LLM-based\\
    \bottomrule
    \end{tabular}
    }
    \label{tab:datasets}
\end{table}

In Petke~et~al.'s study~\cite{petke_patch_overfitting_problem}, the authors report that they examined all generated plausible patches to determine their correctness based on four assessments: syntax comparison with developer-written patches, comparison with patches from previous work, automated dynamic analysis with extra tests, and manual analysis by two independent reviewers for leftover unlabelled patches. Additionally, they report that the assessors have 15+ years of Java development experience. Regarding the RepairLLaMA dataset~\cite{repairllama2023}, the authors report that the first two authors independently labeled all plausible but not AST/Exact match patches and the third author helped resolve any disagreements (which happened in only 4.18\% of cases).

We analyse the two datasets separately for several reasons.
Each comes from a different experimental setup, albeit both reflect practical scenarios in which we expect to observe a natural distribution of correct-to-overfitting patches as generated by the targeted APR tools. 
In Petke~et~al.'s study, the APR tools were allowed to generate patches for up to an 8-hour time limit, to give each tool an equal chance to generate a correct patch.
Each patch is marked with a timestamp, allowing for analysis of any time budget up to the 8-hour mark. 
RepairLLaMA, on the other hand, requires a one-time cost for fine-tuning before use.
Once this step is complete, RepairLLaMA will typically generate up to 20 patches per bug. 
The time cost from the practitioner's perspective is thus LLM-inference (usually in seconds or minutes) plus test execution for up to 20 patches. 
We argue that these datasets demonstrate a realistic upper bound in terms of the number of plausible patches that would be generated for a bug in practice, as it has been shown that roughly a fifth of software engineers are willing to review up to 10 patches per bug~\cite{aprtrust}, while APR can also be run overnight~\cite{JanusManager}.
Moreover, both datasets contain patches generated for Java bugs from Defects4J~\cite{defects4j}, albeit different versions.
To allow for cross-dataset comparisons, in the RepairLLaMA dataset, we filter out patches for bugs that were not attempted by the APR tools in Petke et al.'s study~\cite{petke_patch_overfitting_problem}.
We also performed syntactic patch deduplication, including whitespace consideration. 
Overall, we used \patchesPetke{} unique labelled patches (covering \bugsPetke{} bugs) from the Petke et al.'s work, and \patchesSilva{} unique labelled patches (covering \bugsSilva{} bugs) from Silva et al.'s work. In the remainder of the paper, we will refer to the first dataset as the \textit{\dataOne} one (since it was generated by traditional APR tooling) and to the second one as the \textit{\dataTwo} dataset.

\begin{table}[t]
    \centering
    \caption{POD tools evaluated in this study. We select two tools representing state-of-the-art for each POD category.}
    \resizebox{0.95\columnwidth}{!}{
    \begin{tabular}{lll}
      \toprule
      \textbf{Tool} & \textbf{Type} & \textbf{Summary} \\
      \midrule
      MIPI~\cite{MIPI} & Static & Static analysis for overfitting anti-patterns. \\
      Yang et~al.~\cite{Entropy_delta} & Static & Uses code ``naturalness''/static cues for correctness. \\
      Invalidator~\cite{Invalidator} & Dynamic & Executes extra tests to expose overfitting behavior. \\
      FIXCHECK~\cite{FIXCHECK} & Dynamic & Runtime checks/invariants to reject overfits. \\
      LLM4PatchCorrect~\cite{LLM4PatchCorrect} & Learning & LLM-based classifier of patch correctness. \\
      Tian et al.~\cite{DL4Patch} & Learning & Supervised model trained on labeled patches. \\
      \bottomrule
    \end{tabular}
    }
    \label{tab:tools}
\end{table}

\subsection{Patch Overfitting Detection Tools}\label{sec:tool-selection}
We selected \totDetectors{} POD tools, listed in Table~\ref{tab:tools}, for our empirical study. 
To ensure balanced coverage, we include two tools from each general overfitting detection methodology (Section~\ref{sec:relwork}): 
MIPI \cite{MIPI} and the tool by Yang et al. \cite{Entropy_delta} (static analysis); 
Invalidator \cite{Invalidator} and FIXCHECK \cite{FIXCHECK} (dynamic analysis);
LLM4PatchCorrect \cite{LLM4PatchCorrect}, and the tool by Tian et al. \cite{DL4Patch} (learning-based).

Our selection criteria required POD tools to \textcircled{1} be fully automatic, \textcircled{2} be APR tool agnostic, \textcircled{3} have publicly available source code, and \textcircled{4} have clearly documented replication packages. Finally, more recent proposals were prioritised to reflect current advancements in the field.
For example, we have identified LLM4PatchCorrect~\cite{LLM4PatchCorrect} as representative of the latest LLM-based approaches, as it was shown to outperform earlier learning-based approaches such as ODS~\cite{ods}.
We also exclude from our analysis POD tools that rely on an additional oracle~\cite{fei-2025-pca}, such as those which require human-written (\emph{ground-truth}) patches, e.g., RGT~\cite{ye-et-al-dataset} or DiffTGen~\cite{DiffTGen}.

Most of the tools evaluated in our study required additional information beyond the raw textual representation of patches. In cases where explicit instructions or code to generate the required data were missing or incomplete, the procedures described in each tool's respective publication were closely followed, and customised scripts and minor adjustments were implemented as detailed in our replication package~\repo. 

Aside from the collection and formatting of supplementary information, no significant modifications were made to the original source code of the selected tools. The main adjustments consisted of measuring execution time and capturing each tool's prediction outcomes in a consistently structured CSV file for comparative analysis across all experiments.

\subsection{Computational Resources}\label{sec:computation-specs}
All experiments were performed on Ubuntu 22.04, using an Intel i9-12900K CPU, 32GB of RAM, and a NVIDIA RTX 3090 GPU with 24GB of VRAM.

\section{Results and Discussion}
\label{sec:results}

The \totDetectors{} \toolAcronym{} tools were able to analyse patches with different degrees of success.
MIPI was the least performing as it failed to process 425 out of \patchesPetke{} patches in the \dataOne{} dataset, and 47 out of \patchesSilva{} \dataTwo{} patches, crashing with a  {\small \texttt{javaparser.ParseProblemException}} on those instances.
Both dynamic tools were unable to process the same 21 patches from the \textit{\dataOne{}} dataset, and Invalidator was unable to process one \dataTwo{} patch. In these cases, the dynamic tools' analyses exceeded the configured timeout limits.
The other static tool (Yang et al.) and the learning-based tools successfully processed both complete datasets.
To compare all tools on a level playing field, we excluded MIPI from the analysis of RQ1 and RQ3 due to its poor performance and evaluated the remaining tools on the common set of patches for which all tools were successful, namely 798 and 169 patches from \textit{\dataOne} and \textit{\dataTwo} datasets, respectively.

\subsection{RQ1: Effectiveness of Patch Overfitting Detectors}

\begin{figure}[!t]
\centering

\subfloat{%
\includegraphics[width=0.85\columnwidth]{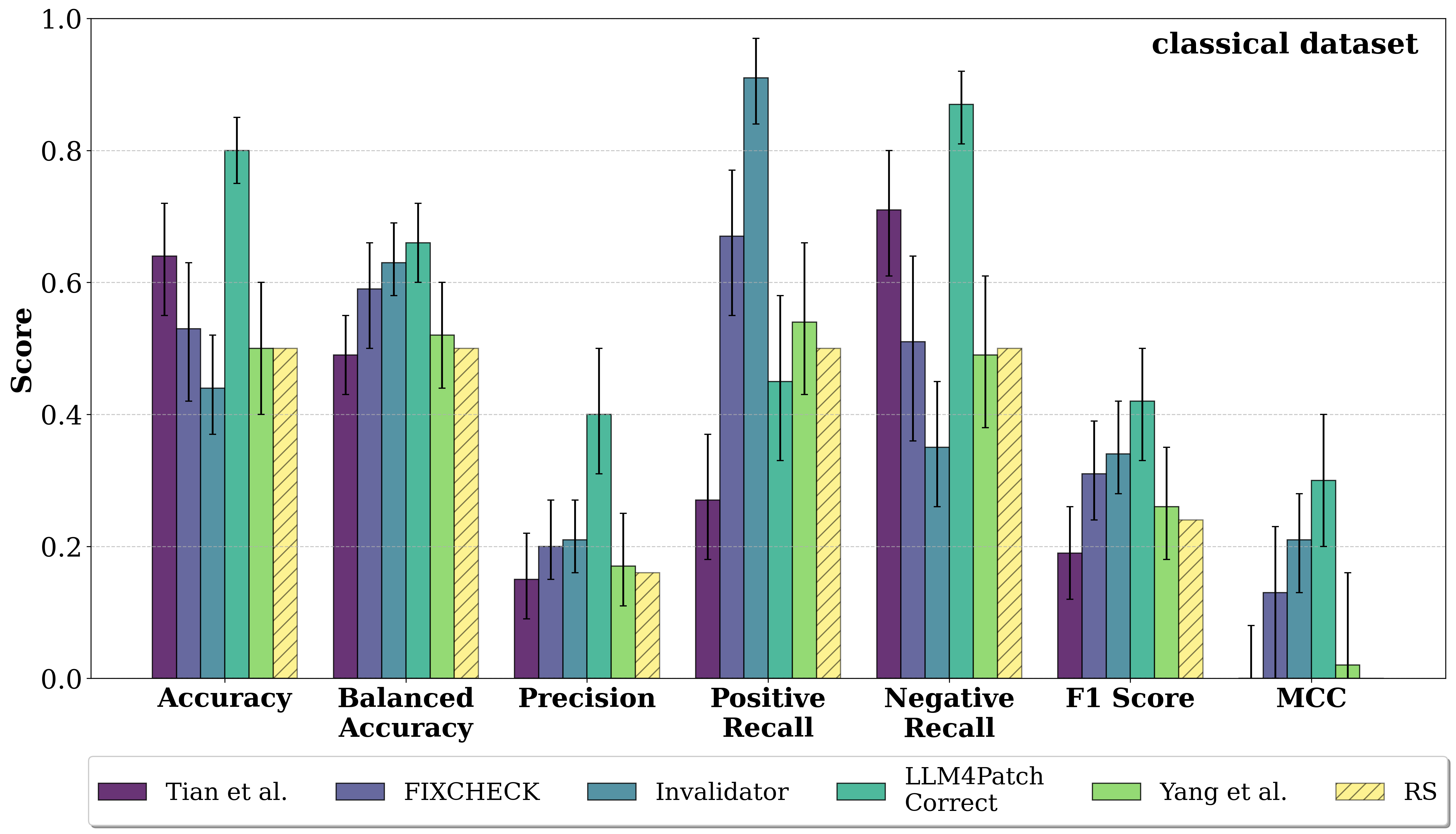}%
}

\subfloat{%
\includegraphics[width=0.85\columnwidth]{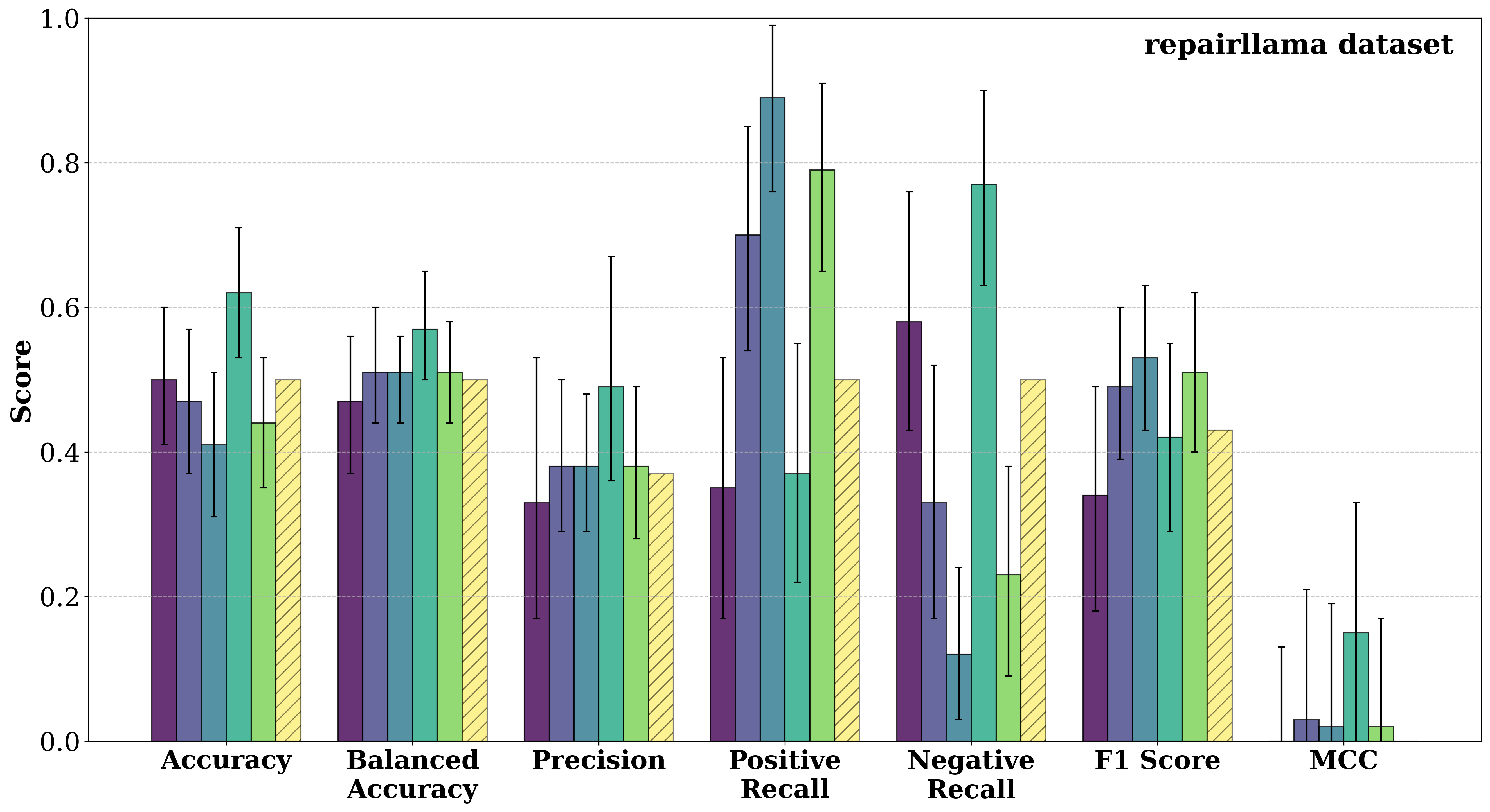}%
}
\caption{RQ1. Performance of 5 POD tools and RS baseline on the \dataOne{} (top) and \dataTwo{} (bottom) datasets.}
\label{fig:ovr-classification-metrics}
\end{figure}

Fig.~\ref{fig:ovr-classification-metrics} contrasts the 5 \toolAcronym{} tools with the RS baseline on seven standard metrics computed over the \dataOne{} (top fig.) and \dataTwo{} (bottom fig.) datasets.
Each bar represents the point estimate, and whiskers indicate 95\% cluster-aware bootstrap confidence intervals.
Specifically, we treated the patches generated by each APR tool and bug combination as a single cluster. 
This resulted in 242 unique clusters for the \dataOne{} dataset and 50 for the \dataTwo{} dataset, with cluster sizes ranging from 1 to 61 patches. 
The median cluster contains a single patch for the \dataOne{} dataset and three patches for the \dataTwo{} dataset, while the mean is 3.3 and 3.4 patches (standard dev. $\approx 6.3$ and $\approx 2.18$), accordingly. 
We then drew 1,000 bootstrap replicates, resampling these clusters with replacement and re-computing the metric on the pooled patches, to capture inter-cluster variability. 
The confidence intervals represent the 2.5\textsuperscript{th} and 97.5\textsuperscript{th} percentiles.\looseness=-1 

\subsubsection{Accuracy and Balanced Accuracy}
For the \dataOne{} dataset, LLM4PatchCorrect achieves the highest performance across both metrics (Accuracy = 0.80, Balanced Accuracy = 0.66) and is the only tool to outperform RS in both metrics simultaneously. Tian et al. ranks second in raw accuracy, with a score of 0.64 and confidence intervals above the baseline. The remaining tools cannot be distinguished from random guessing in raw accuracy. Despite their low accuracy scores, the two dynamic-based tools exhibit the second- and third-highest balanced accuracy scores (Invalidator = 0.63, FIXCHECK = 0.59), indicating a pronounced strength in detecting the minority class of correct patches.\looseness=-1

For the \dataTwo{} dataset, although the values are lower across the board, LLM4PatchCorrect still comes out on top with values of 0.62 and 0.57 for accuracy and balanced accuracy, respectively. 
However, as the lower bound of the confidence interval for balanced accuracy matches the baseline, it only significantly outperforms the baseline in terms of raw accuracy. 
Suffering from degraded performance, Tian et al. 
now matches RS on raw accuracy. 
Following the same pattern as the \dataOne{} dataset, the remaining tools fall below the baseline on this metric for the \dataTwo{} dataset, making four out of five techniques indistinguishable from random guessing. 
For balanced accuracy, FIXCHECK, Invalidator, and Yang et al. all tied at 0.51; however, none of these significantly outperforms the baseline.

\subsubsection{Precision–Recall Trade-off}
Fig. \ref{fig:ovr-classification-metrics} shows a clear precision-recall trade-off across the tools evaluated. 
For the \dataOne{} dataset, Invalidator achieves the highest positive-class recall of 0.92, successfully identifying most correct patches. However, it suffers significantly from over-predicting correctness, resulting in low precision (0.21) and negative recall (0.35), which is worse than the RS baseline. 
We see this pattern carry over in the \dataTwo{} dataset, where Invalidator leads in positive recall (0.89) at the cost of having by far the weakest negative recall (0.12) and an average precision (0.38) that barely beats the baseline. 
For negative recall, LLM4PatchCorrect excels, identifying 87\% of overfitting patches for the \dataOne{} dataset and 77\% for the \dataTwo{} dataset.
It also leads in precision across both datasets (0.40 and 0.49), although this performance is accompanied by a relatively modest positive recall (0.45 and 0.37).
This pattern of prioritising the detection of overfitting patches at the expense of correct ones is common among the two evaluated learning-based methods.
In comparison, Invalidator discarded far fewer correct patches (8\%), but allowed more overfitting patches to pass than any other approach, failing to identify 65\% of the overfitting patches in the \dataOne{} dataset. 
Dynamic tools performed best in retaining correct patches, with FIXCHECK having the second-highest and only other significant positive recall score among the tools.

Interestingly, the performance of the Yang et al. tool varies significantly across datasets. 
For the \dataOne{} dataset, it exhibited a more balanced recall (0.54 positive; 0.49 negative), but its low precision (0.17) suggests that it flagged many patches as ``correct” that in fact overfit the test suite. 
In our confidence tests, this tool did not outperform the baseline on any metric. 
However, for the \dataTwo{} dataset, we observe a shift, as the Yang et al. tool now ranks 2nd (0.79) in positive recall, surpassing the baseline significantly. 
However, this is accompanied by a significant decline in negative recall (0.23), which falls far below the random baseline, indicating the tool is massively over-predicting correctness. 
This may be because LLM-generated patches tend to have higher naturalness than those generated by classical APR tools~\cite{natural}, which the tool uses to predict overfitting.

Tian et al.'s tool, which shows significant improvement over the baseline only in accuracy and negative recall, demonstrates a strong preference for specificity, correctly identifying 71\% of overfitting patches. 
However, it fails to detect 73\% of correct patches for the \dataOne{} dataset, substantially limiting its practical effectiveness. 

\subsubsection{Harmonic and Correlation Measures}
In the \dataOne{} dataset, the F1-score, which harmonises precision and positive recall, is highest for LLM4PatchCorrect (0.42), followed by Invalidator (0.34). 
Both approaches significantly outperform the RS baseline. 
Notably, the confidence intervals of the other three tools cannot outperform random guessing in F1-score, with the learning-based Tian et al. tool having the lowest score. 
This finding is in contrast with the \dataTwo{} dataset results, where LLM4PatchCorrect's relative drop in positive recall places it in fourth place for F1-score (0.42), while Invalidator and Yang et al. take first and second place with scores of 0.53 and 0.51, respectively. 
For this dataset, none of the confidence intervals of the five tools significantly surpass the baseline.
However, F1-score alone does not capture performance on the negative class. 
MCC, which accounts for all four confusion‐matrix quadrants and is robust under class imbalance, is highest for LLM4PatchCorrect at 0.30 (0.20–0.40) on the \dataOne{} dataset. 
The nearest performer in MCC is Invalidator at 0.21 (0.13–0.29), whilst FIXCHECK partially outperforms the baseline of 0.
On the contrary, Tian et al. and Yang et al. have near‐zero MCC, effectively matching that of random prediction. Overall, we see lower MCC values for the \dataTwo{} dataset, with no tools significantly outperforming the baseline.

\subsubsection{Comparative Classification Performance on Each Class}
\begin{figure}[!t]
\centering

\subfloat{%
\includegraphics[width=0.85\columnwidth]{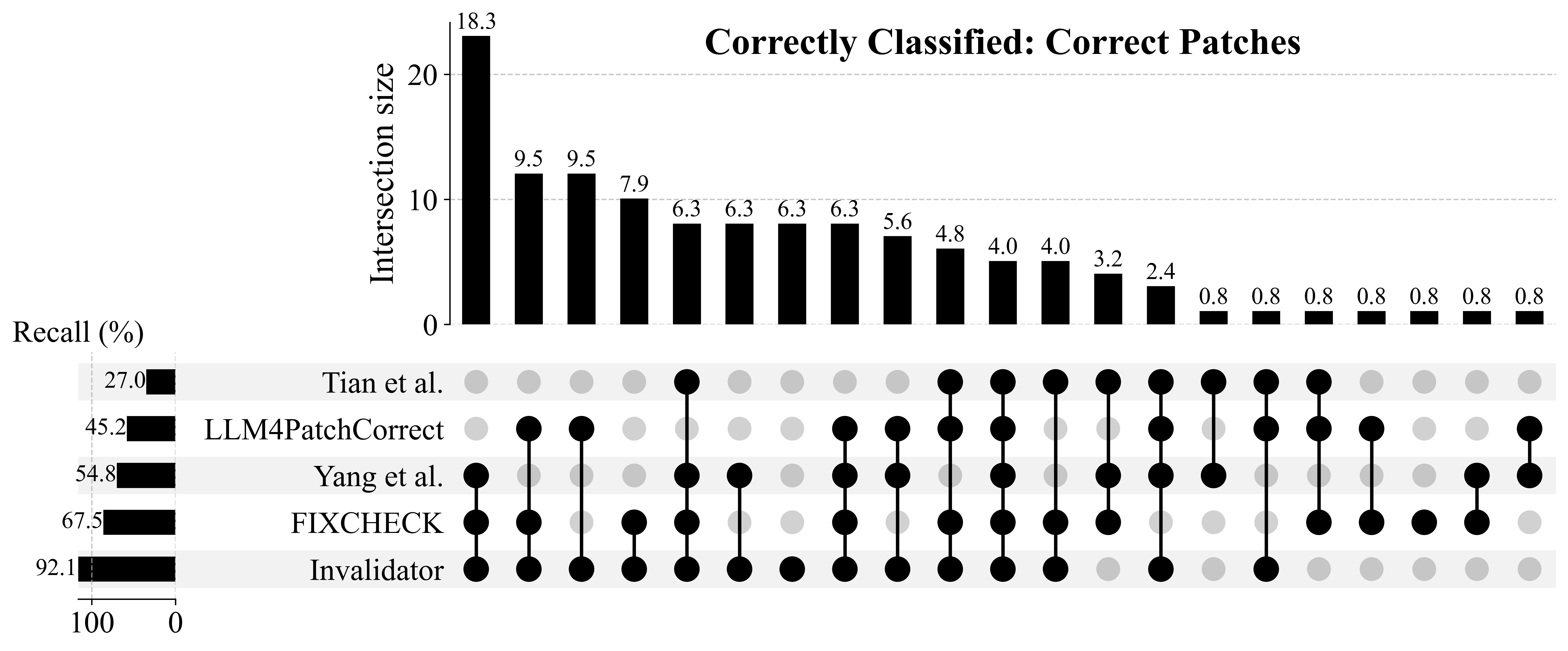}%
}

\subfloat{%
\includegraphics[width=0.85\columnwidth]{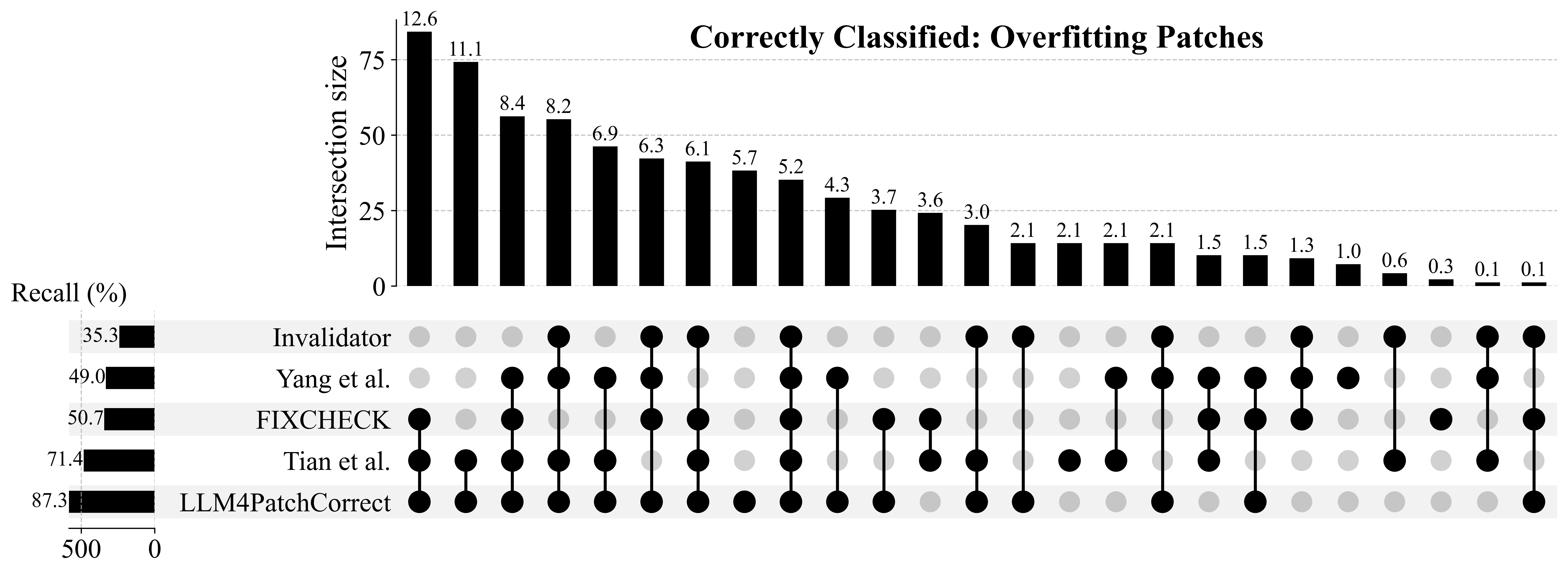}%
}
\caption{RQ1. UpSet plot of Correct Patches (top, 127) and Overfitting Patches (bottom, 671) that were correctly classified by each tool or subsets of tools for the \dataOne{} dataset. The bar charts on the bottom left represent recall for each tool. Each dot corresponds to a set of correctly classified patches. Edges represent intersections, with the top bar chart representing the size of the intersections (\%s on bar tops).}
\label{fig:upset-correct-overfitting}
\end{figure}

The UpSet plots presented in Fig. \ref{fig:upset-correct-overfitting} illustrate the comparative performance of the evaluated tools in correctly identifying unique sets of correct and overfitting patches, respectively, for the \dataOne{} dataset (equivalent UpSet plots for the \dataTwo{} dataset are available in~\repo). 
On top, it visualises the correct classification of the 127 fixes, while the bottom addresses the classification of the 671 overfitting patches. 
The horizontal bars indicate per-tool recall, whereas the vertical bars, arranged in descending order, represent intersection sizes among tool classifications.\looseness=-1

Notably, the tallest intersection bar of 23 patches detected by Invalidator, Yang et al. and FIXCHECK was wrongly classified by the two learning-based approaches.
In fact, the dynamic tools (Invalidator \& FIXCHECK) identified 51 patches that the learning-based approaches (LLM4PatchCorrect \& Tian et al.) missed, whereas the learning-based methods uncovered only two patches that the dynamic approaches overlooked. Overall, dynamic approaches performed best on this metric, especially Invalidator, where the largest intersection it did not contribute to comprised only 4 patches.

Overall, for the class of correct patches, 4 out of 5 tools are able to accurately detect at least 57 patches (45.2\%), with a sizable overlap in detection performance. Despite this, the fourth biggest subset of 8 patches could only be detected by Invalidator, with the only other tool-exclusive subset belonging to FIXCHECK (1 patch). This highlights potential complementary strengths that an ensemble could exploit. 
There are no tool-exclusive subsets for the \dataTwo{} results. For the correct patches in the \dataTwo{} dataset, we observed similar patterns, although the positions of Yang et al. and FIXCHECK are swapped.\looseness=-1

For overfitting patches, 4 out of 5 tools are able to accurately detect at least 329 patches in the \dataOne{} dataset (49\%). Again, there was a significant overlap in detection capabilities, at a similar rate to the class of correct patches.
Interestingly, the relative order of performance among types of detection tools is largely reversed compared to their detection of correct patches. Here, learning-based tools markedly outperform other approaches. In particular, 18.9\% of overfitting patches (127) are exclusively recognised by the learning-based methods, reinforcing the previously observed pattern that these methods are the most effective at discarding invalid repairs. This finding is even more extreme in the \dataTwo{} results, where 48.6\% of overfitting patches were only correctly classified by the learning-based approaches.

The set-intersection view in Figure~\ref{fig:upset-correct-overfitting} reveals a result which is not visible from class-level recall alone. Nearly every correct patch recognised by either learning-based method is already found by at least one dynamic analyser, and the overfitting patches caught by dynamic analysers are overwhelmingly those that the learning-based methods also reject. Consequently, in the class where each category is comparatively weaker, the fraction of patches it detects that are not already identified by the other category is small. These observations also extend to the \dataTwo{} dataset.

\subsubsection{Bug‑Level Discrimination}
\begin{figure}[!t]
    \centering
\includegraphics[width=0.85\linewidth]{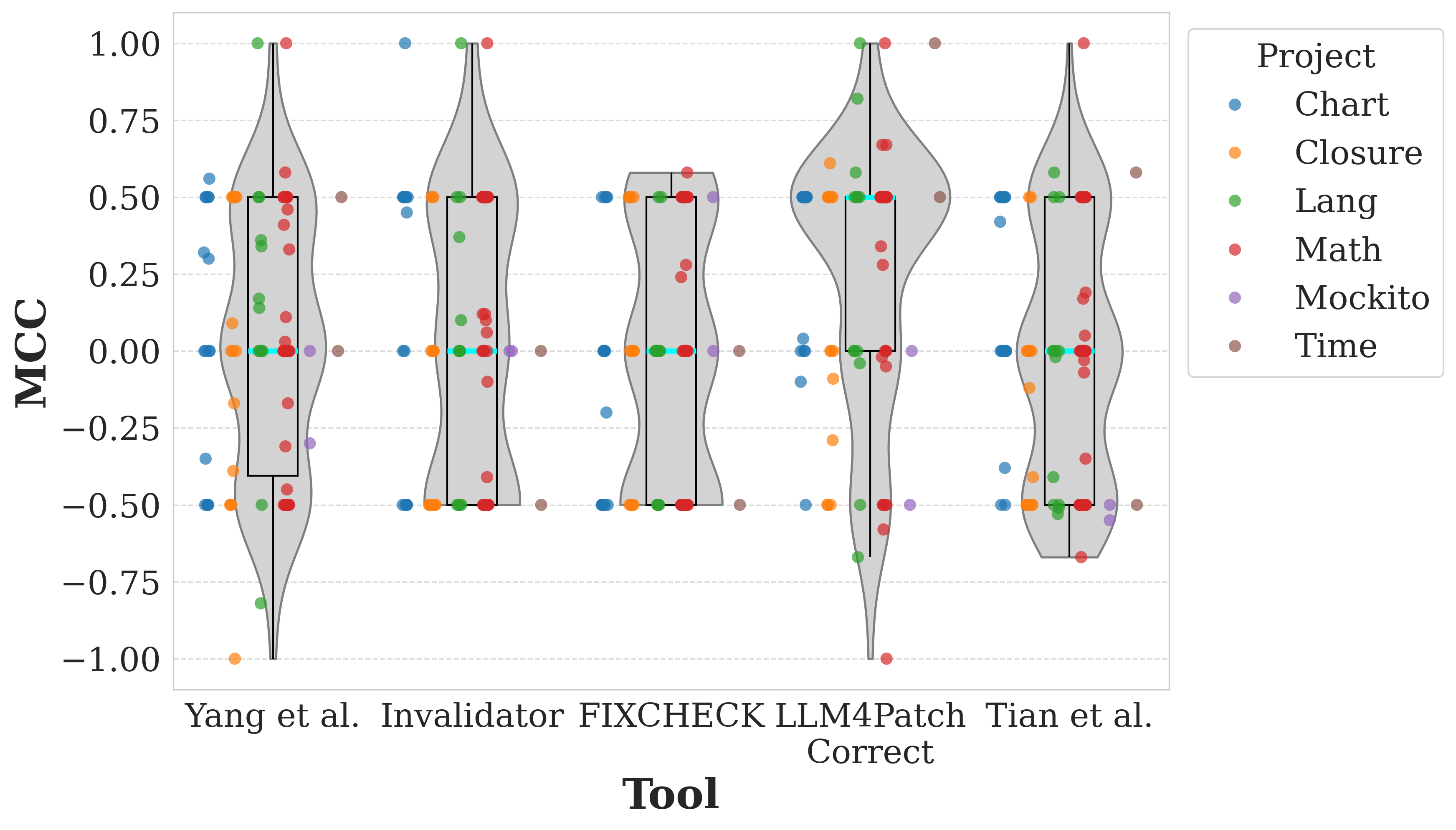}
    \caption{RQ1. MCC Scores on Bug-Level Predictions per Tool and per Project for the \dataOne{} dataset.}
    \label{fig:mcc-violin}
\end{figure}

Figure \ref{fig:mcc-violin} shows, for each detection tool, a violin plot of the distribution of bug‑level MCC scores on the \dataOne{} dataset. Each dot is one Defects4J bug, colour‑coded by its corresponding project. 
The violin’s shape reveals both the central tendency and variance of each detector’s discriminative power across bugs and projects. The cyan line inside each boxplot per violin indicates the median.

LLM4PatchCorrect is clearly dominant at  bug-level granularity. Its violin is centred on a median MCC of 0.50, the only strictly positive median among all methods, and its inter‑quartile range is the narrowest of the group. The width of the violin around 0.4–0.6 shows that a large mass of bugs are classified moderately well, whereas the slim tail at 0 or below shows that LLM4PatchCorrect does not degrade to chance or systematic misclassifications as much as other tools.

By contrast, the remaining detectors all exhibit medians at exactly 0. This value implies that, for the majority of bugs, these tools behave no better than random guessing; moreover, their violins are visibly wider in the $\leq 0$ domain, indicating frequent systematic errors on certain bugs. The \dataTwo{} results paint a bleaker picture: here, all five tools exhibit medians of 0. We also see the widest point for each violin around this median, suggesting that no tool demonstrates significant strength in classifying the LLM-generated patches at a bug-level granularity.

Perfect bug classification is rare, with Invalidator and LLM4PatchCorrect each achieving an MCC score of 1 on three bugs of the \dataOne{} dataset; the highest among all tools. At the same time, perfect misclassification is even less common: Yang et al. and LLM4PatchCorrect achieve a -1 MCC score on a bug each. We see similar numbers in our \dataTwo{} results, where Invalidator and Tian et al. classified patches for two bugs perfectly, and Yang et al. perfectly misclassified patches for one bug.

In terms of MCC performance by project, Figure \ref{fig:mcc-violin} illustrates that MCC scores for individual projects are scattered widely across the full range of values. Consequently, it is challenging to visually identify consistent clusters or clear patterns of high or low MCC performance for any specific project. To better understand and clarify these trends, a second analysis was conducted, grouping MCC scores explicitly by project rather than by detection tool. 
The resulting project-level box plots for both datasets are shown in our replication package~\repo.
In our \dataOne{} dataset results, we observe that every project exhibits a median MCC of precisely 0, confirming the absence of any systematic advantage for or against a particular project. In other words, once tool effects are factored out, no project emerges as reliably predictable at the bug level.
In addition, five of the six projects have 3rd quartiles at 0.50, whereas Mockito tops out at 0.00. Taken with its relatively low 1st quartile (-0.45), this suggests that detectors struggle disproportionately with the two Mockito bugs, rarely achieving even moderate discrimination. 
Chart is the most forgiving project: the minimum MCC is -0.50 and its first quartile is only -0.24, indicating that outright misclassification is comparatively uncommon. By contrast, Closure and Math reach the theoretical floor of -1, implying certain bugs in these projects systematically confuse most detectors.\looseness=-1

We observe similar patterns in the \dataTwo{} MCC-per-bug values. All projects have a median MCC of 0 except for Time, which has a median of -0.5 (though only one bug from this project is included in the dataset). The most notable difference is that 3rd-quartile values are generally lower, with only Closure and Math reaching an MCC of 0.5. 
Chart is the only project where a tool completely misclassified the patches for a bug in this dataset.

Collectively, these findings reinforce the conclusion drawn from the violin plots: variation in bug-level performance is dominated by individual bugs rather than the project they belong to. This is useful when extrapolating to unseen codebases or benchmarks outside Defects4J.

\subsubsection{APR‑Tool Sensitivity} 
\begin{figure}[!t]
    \centering
\includegraphics[width=0.85\linewidth]{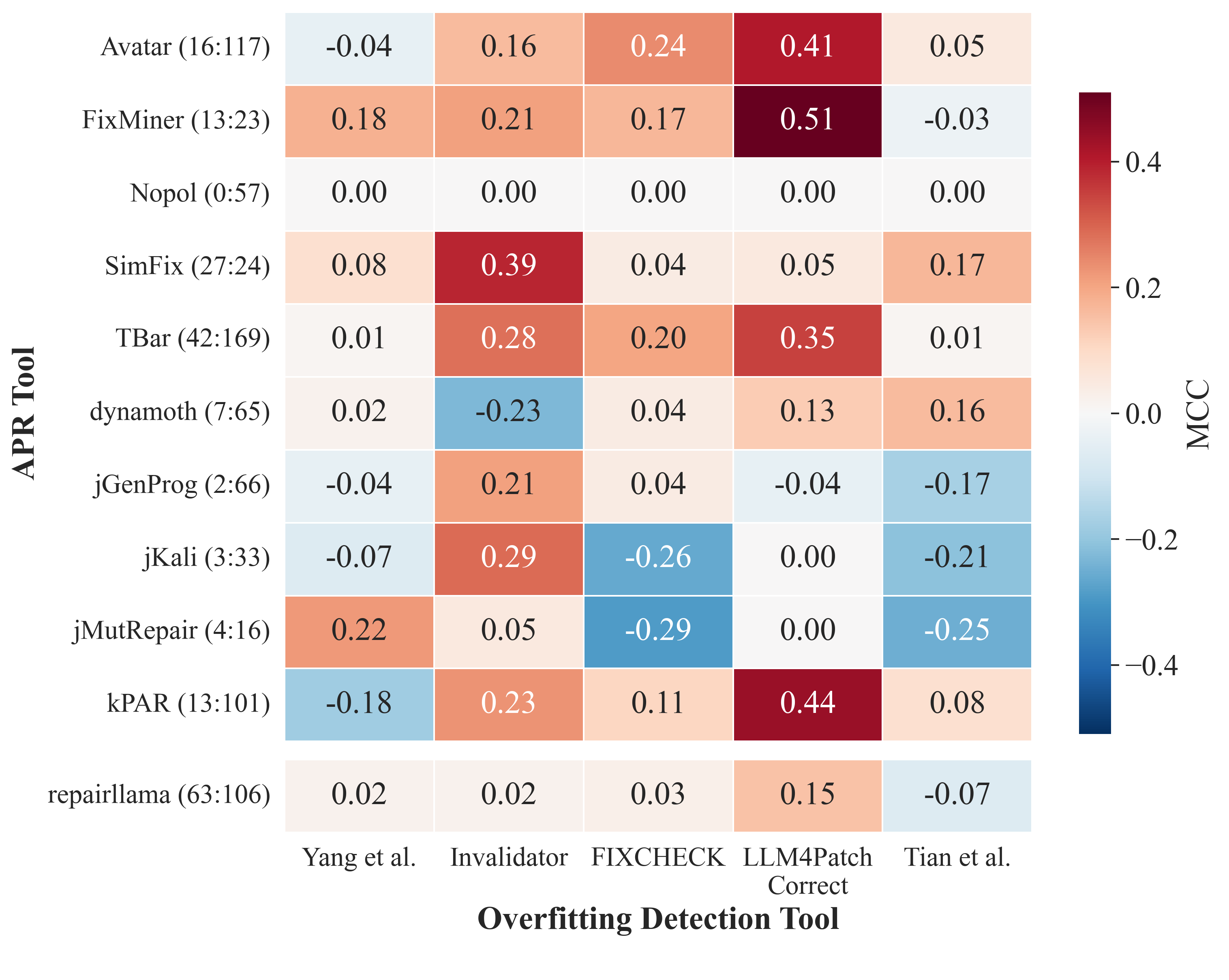}
    \caption{RQ1. Matrix of Average MCC Scores for Patches Generated by APR Tools. The numbers in brackets after the APR tool name give the class balance (\#correct : \#overfitting).}
    \label{fig:mcc-apr}
\end{figure}

We now perform a complementary analysis to understand how detection tool performance varies with the type of APR output. To do so, we aggregate MCC scores by APR tool. The matrix in Figure \ref{fig:mcc-apr} averages the bug‑level MCC of every detector within the patches emitted by a single APR tool for the \dataOne{} and \dataTwo{} datasets.

LLM4PatchCorrect again leads this metric, securing the best or joint‑best MCC on 5 of the 10 \dataOne{} APR tools and also for \dataTwo{}.
Overall, no strong trends emerge in terms of APR tools that systematically produce more easily classifiable patches, or in detectors that maintain uniform performance irrespective of APR tool characteristics.
It appears that there is only a weak correlation between the type of APR used and the POD technique. Arguably, it seems that LLM4PatchCorrect performs better for tools that use template-based approaches (Avatar, FixMiner, TBar, kPAR).
However, these are only initial observations that would require a different experimental setup to investigate further.

\subsection{RQ2: Efficiency of Patch Overfitting Detectors}

\begin{table}[!t]
\caption{RQ2. Total execution time (h:min:sec) and time (in seconds) per successfully classified patch for each POD tool on the \dataOne{} and \dataTwo{} datasets.}
\begin{center}
{\setlength\tabcolsep{3pt}
\resizebox{0.95\columnwidth}{!}{
\begin{tabular}{llrr}
\toprule
& & \multicolumn{2}{c}{\textbf{Execution Time (\# Classified, Seconds/Patch)}} \\
\cmidrule{3-4}
\textbf{POD Tool} & \textbf{Type} & \multicolumn{1}{c}{\textbf{\dataOne{} (819 Patches)}} & \multicolumn{1}{c}{\textbf{\dataTwo{} (170 Patches)}} \\
\midrule
Tian et al. \cite{DL4Patch} & Learning & 00:11:51 (819, 0.87) & 00:02:28 (170, 0.87) \\
MIPI \cite{MIPI} & Static & 00:06:30 (373, 1.05) & 00:01:51 (123, 0.90) \\
Yang et al. \cite{Entropy_delta} & Static & 01:38:49 (819, 7.24) & 00:23:18 (170, 8.22) \\
LLM4PatchCorrect \cite{LLM4PatchCorrect} & Learning & 01:52:14 (819, 8.22) & 00:27:43 (170, 9.78) \\
FIXCHECK \cite{FIXCHECK} & Dynamic & 15:26:32 (798, 69.66) & 02:50:49 (170, 60.29) \\
Invalidator \cite{Invalidator} & Dynamic & 41:16:49 (798, 186.23) & 15:17:57 (169, 325.90) \\
\bottomrule
\end{tabular}
}
}
\label{tab:tool_exec_time}
\end{center}
\end{table}

Table~\ref{tab:tool_exec_time} shows the overall execution time of running each POD tool, the number of patches each tool successfully classified, and the execution time per patch for each tool. 
Overall, we observe that static and learning-based tools are faster than dynamic tools.
While it could be thought that greater computational effort (i.e., higher time per patch) should yield improved detection of overfitting patches, our results show that is not always the case.
In fact, the classification results discussed in RQ1 should be interpreted in light of runtime findings. 
The dynamic tools Invalidator and FIXCHECK demonstrate high positive recall, that is, strong performance in retaining correct patches, but at a higher computational cost overall. 
In contrast, LLM4PatchCorrect strikes a practical balance, offering superiority compared to the baseline in all metrics except positive recall on the \dataOne{} dataset, while remaining computationally efficient. This performance-to-efficiency trade-off does not extend to the \dataTwo{} dataset, where LLM4PatchCorrect is only significantly superior to the RS baseline in two metrics (accuracy and negative recall). Overall, depending on the deployment context and the number of patches required to be reviewed, any of the tools we consider could be viable for use in practice in terms of runtime.

\subsection{RQ3: Patch Overfitting Detectors Vs. Baselines}

\subsubsection{RS Baseline}\label{sec:rs-baseline-results}
Table \ref{tab:rs-baseline-table} reports, for every bug that contains at least one correct and one overfitting patch in the \dataOne{} dataset, how many patches a developer must examine before encountering a correct one. 
Overall, the \dataOne{} APR tools produced at least one correct patch for 45 out of the original 92 plausibly patched bugs. For 16 of these bugs, all generated patches were correct. As such, these have been omitted from the table since developers would only need to inspect a single patch to guarantee finding a correct one.
However, it is still worth noting that for 13 of these 16 bugs, at least one POD tool classified all patches as overfitting, discarding all correct patches the developer could review.
The complete table including the omitted rows, and equivalent table for the \dataTwo{} dataset (with rows for 42 out of the original 50 plausibly patched bugs, of which 12 did not have any overfitting patches) are included in our replication package~\repo. 
RS-85 and RS-95 give the theoretical effort when patches are inspected at random until the probability of encountering a correct one is $\geq$85\% or $\geq$95\%, respectively (Section~\ref{sec:rs-baseline}). 
Therefore, a higher confidence level (RS-95) requires \textbf{more} inspections than a lower one (RS-85).
For a POD tool, the number of patches to examine is simply all FP + 1 (all overfitting patches the tool failed to discard, plus the first candidate patch that is actually correct). To illustrate this with an example, consider a scenario where there are 10 patches for a given bug, of which 2 are correct and 8 are overfitting. If a POD tool accurately labels the correct patches (2 TP) but also misclassifies 3 overfitting patches as correct (3 FP), we say the developer needs to review 4 patches (3 FP + 1 TP) before encountering a correct patch. When a tool misses all correct patches, we record N/A.
\begin{table}[!t]
\begin{center}
\caption{RQ3. Number of patches that must be inspected to find a correct patch for each bug where at least one correct AND one overfitting patch was generated by the \dataOne{} APR tools, across POD tools and random selection (RS-85, RS-95). `N/A' indicates that a tool was unable to classify any true positives. Bold values signify the best performance per bug. The ratio of correct-to-overfitting patches generated by the APR tools is provided next to the bug name.
}
{\setlength\tabcolsep{5pt}
\resizebox{0.85\columnwidth}{!}{
\begin{tabular}{lrrrrrrr}
\toprule
\textbf{Bug} & \textbf{Yan.} & \textbf{Inv.} & \textbf{FIX.} & \textbf{LLM.} & \textbf{Tia.} & \textbf{\makecell{RS\\-85}} & \textbf{\makecell{RS\\-95}} \\
\midrule
Chart-1 \scriptsize{(8:21)} & \textbf{2} & 22 & N/A & 8 & N/A & 6 & 8 \\
Chart-4 \scriptsize{(5:1)} & \textbf{1} & \textbf{1} & 2 & N/A & N/A & 2 & 2 \\
Chart-7 \scriptsize{(1:34)} & 3 & \textbf{1} & N/A & N/A & N/A & 30 & 34 \\
Chart-9 \scriptsize{(1:10)} & N/A & 11 & 11 & N/A & \textbf{4} & 10 & 11 \\
Closure-2 \scriptsize{(1:2)} & N/A & \textbf{3} & \textbf{3} & \textbf{3} & N/A & \textbf{3} & \textbf{3} \\
Closure-46 \scriptsize{(4:6)} & 5 & 7 & 7 & \textbf{1} & N/A & 4 & 5 \\
Closure-57 \scriptsize{(2:2)} & \textbf{3} & \textbf{3} & \textbf{3} & N/A & N/A & \textbf{3} & \textbf{3} \\
Closure-62 \scriptsize{(4:14)} & 10 & 15 & 15 & N/A & 10 & \textbf{7} & 9 \\
Closure-126 \scriptsize{(5:14)} & 10 & 15 & N/A & \textbf{5} & N/A & 6 & 8 \\
Lang-10 \scriptsize{(3:7)} & \textbf{3} & 8 & 8 & N/A & N/A & 5 & 6 \\
Lang-33 \scriptsize{(3:2)} & \textbf{2} & 3 & 3 & 3 & 3 & \textbf{2} & 3 \\
Lang-43 \scriptsize{(2:2)} & 2 & 3 & 3 & \textbf{1} & 2 & 3 & 3 \\
Lang-55 \scriptsize{(5:8)} & 7 & \textbf{1} & 9 & N/A & 9 & 4 & 5 \\
Lang-57 \scriptsize{(5:5)} & N/A & 6 & N/A & \textbf{2} & N/A & 3 & 4 \\
Lang-58 \scriptsize{(1:53)} & \textbf{27} & 36 & 54 & N/A & N/A & 46 & 52 \\
Lang-59 \scriptsize{(4:25)} & \textbf{1} & \textbf{1} & N/A & \textbf{1} & N/A & 11 & 15 \\
Math-30 \scriptsize{(4:6)} & 5 & 6 & 7 & \textbf{3} & N/A & 4 & 5 \\
Math-33 \scriptsize{(3:97)} & 96 & 68 & \textbf{27} & N/A & N/A & 47 & 63 \\
Math-50 \scriptsize{(9:73)} & N/A & 74 & 48 & 34 & 20 & \textbf{15} & 23 \\
Math-53 \scriptsize{(1:2)} & \textbf{1} & 3 & 3 & N/A & \textbf{1} & 3 & 3 \\
Math-57 \scriptsize{(2:2)} & \textbf{2} & 3 & \textbf{2} & \textbf{2} & N/A & 3 & 3 \\
Math-58 \scriptsize{(1:18)} & N/A & 18 & N/A & \textbf{1} & N/A & 17 & 19 \\
Math-70 \scriptsize{(3:2)} & \textbf{1} & 3 & 3 & \textbf{1} & 2 & 2 & 3 \\
Math-79 \scriptsize{(1:3)} & \textbf{4} & \textbf{4} & N/A & N/A & N/A & \textbf{4} & \textbf{4} \\
Math-80 \scriptsize{(1:47)} & 8 & \textbf{1} & 48 & N/A & N/A & 41 & 46 \\
Math-82 \scriptsize{(3:17)} & 7 & 17 & N/A & \textbf{6} & N/A & 9 & 12 \\
Math-85 \scriptsize{(4:51)} & 29 & 28 & N/A & N/A & \textbf{4} & 21 & 29 \\
Mockito-29 \scriptsize{(2:5)} & 5 & 6 & 6 & N/A & N/A & \textbf{4} & 5 \\
Time-7 \scriptsize{(4:4)} & 4 & 5 & N/A & \textbf{1} & 3 & 3 & 4 \\
\bottomrule
\end{tabular}
}
}
\label{tab:rs-baseline-table}
\end{center}
\end{table}

Comparing each POD tool individually against the RS baselines demonstrates the strength of random selection as a benchmark. At both 85\% (RS-85) and 95\% (RS-95) thresholds, random selection consistently equals or outperforms individual detectors on the majority of bugs. Specifically, RS-95 matches or surpasses Yang et al. on 32 out of 45 bugs (71\%), LLM4PatchCorrect on 34 (76\%), Invalidator on 37 (82\%), Tian et al. on 38 (84\%), and FIXCHECK on 43 (96\%). The less thorough RS-85 further improves these results, matching or outperforming Yang et al. on 33 bugs (73\%), LLM4PatchCorrect on 34 (76\%), Invalidator on 39 (87\%), Tian et al. on 41 (91\%), and FIXCHECK on 43 (96\%).

Evaluating the overall performance, RS-85 achieves the best score across all five tools on 23 bugs (51\%), while RS-95 matches or beats all tools on 21 bugs (47\%). 
The two learning-based detectors, LLM4PatchCorrect and Tian et al., primarily underperform because they frequently miss correct patches entirely, with no correct patch detected for 21 and 29 bugs, respectively. Invalidator misses all fixes only once (for a bug where all generated patches were correct), yet its frequent inclusion of overfitting patches causes it to fall short against both RS benchmarks for most bugs.

\begin{table}[!t]
\caption{RQ3. Median and Mean Number of Patches to Inspect (N/As from Table~\ref{tab:rs-baseline-table}  filled with total patches per bug)}
\begin{center}
\resizebox{0.85\columnwidth}{!}{
\begin{tabular}{lrrrr}
\toprule
& \multicolumn{2}{c}{\dataOne{} (45 bugs)} & \multicolumn{2}{c}{\dataTwo{} (42 bugs)} \\
\textbf{Tool} & \textbf{Median} & \textbf{Mean} & \textbf{Median} & \textbf{Mean} \\
\midrule
Yang et al.            & 3.0  & 8.53    & 2.0   & 3.05  \\
Invalidator            & 3.0  & 8.62    & 2.0   & 2.90  \\
FIXCHECK               & 3.0  & 11.29   & 2.0   & 3.02  \\
LLM4PatchCorrect       & 3.0  & 10.36   & 2.5   & 3.12  \\
Tian et al.            & 4.0  & 11.31   & 2.0   & 3.07  \\
RS-85                  & 3.0  & 7.42    & 2.0   & 2.69  \\
RS-95                  & 3.0  & 9.02    & 2.0   & 2.93  \\
\bottomrule
\end{tabular}
}
\label{tab:median-mean-inspection}
\end{center}
\end{table}

Table~\ref{tab:median-mean-inspection} deepens this comparison by summarising, across the 45 bugs in the \dataOne{} dataset and 42 bugs in the \dataTwo{} dataset, the typical (median) and expected (mean) number of patches a developer must inspect before encountering a correct patch identified by either the POD tools or the RS-85 and RS-95 baselines. For this table, in cases where a POD tool fails to identify any of the correct patches for a bug (instances marked N/A in Table~\ref{tab:rs-baseline-table}), we add the full pool of candidate patches to the average calculations. This is to faithfully reflect the worst-case workload where the developer would need to review all patches to find a correct one after the POD tool failed to do so.

Across both datasets, no tool outperforms even RS-95 in terms of medians. Instead, 4 out of 5 tools match the RS-95 median on both occasions, with Tian et al. falling behind in the \dataOne{} dataset results and LLM4PatchCorrect in the \dataTwo{} dataset results. Consequently, for a representative bug, these detectors appear no better than chance. The outcomes for means are similar, with 3 out of 5 tools performing worse than RS-95 on the \dataOne{} dataset. For \dataTwo{} patches, Invalidator is the only tool that beats RS-95 on mean, while all others fall short. In particular, the learning-based detectors (LLM4PatchCorrect \& Tian et al.), although precise when they succeed, often discard every correct patch; those null outputs outbalance their detection performance on other bugs, requiring manual review for the full set of generated patches.\looseness=-1

Interestingly, for more than half of the bugs in each dataset, a developer needs to inspect
\emph{a maximum of three} patches chosen at random to reach 85\% confidence of finding a correct patch (for \dataOne{} dataset: number of rows in Table \ref{tab:rs-baseline-table} where the RS-85 value is $\leq 3$ plus the 16 bugs where all patches were correct), highlighting the challenging bar the RS baselines set when patch count per bug is low. Future evaluations should thus consider including this metric to better track progress relative to this robust baseline.

Since a developer is more likely to run APR tools for a shorter time period than the full eight-hour budget used in our dataset to collect candidate patches, we repeated the analysis using only patches generated in the first hour of each \dataOne{} APR tool’s execution. With this time budget, 39 bugs were patched, as opposed to 45 with the full budget. However, the resulting pattern is unchanged: the minimal-competence bar of RS-95 still beats every detector on a majority of the 39 bugs, surpassing Yang et al. on 25 bugs (64\%), LLM4PatchCorrect on 30 (77\%), Tian et al. on 32 (82\%), Invalidator on 34 (87\%), and FIXCHECK on 38 (97\%). Mean inspection counts decrease for all tools compared with the eight-hour setting, and there is still a considerable amount of failed detections (N/As). The figures for these results can be found in our repository \repo.

\begin{figure}[t]
    \centering
    \includegraphics[width=0.95\columnwidth]{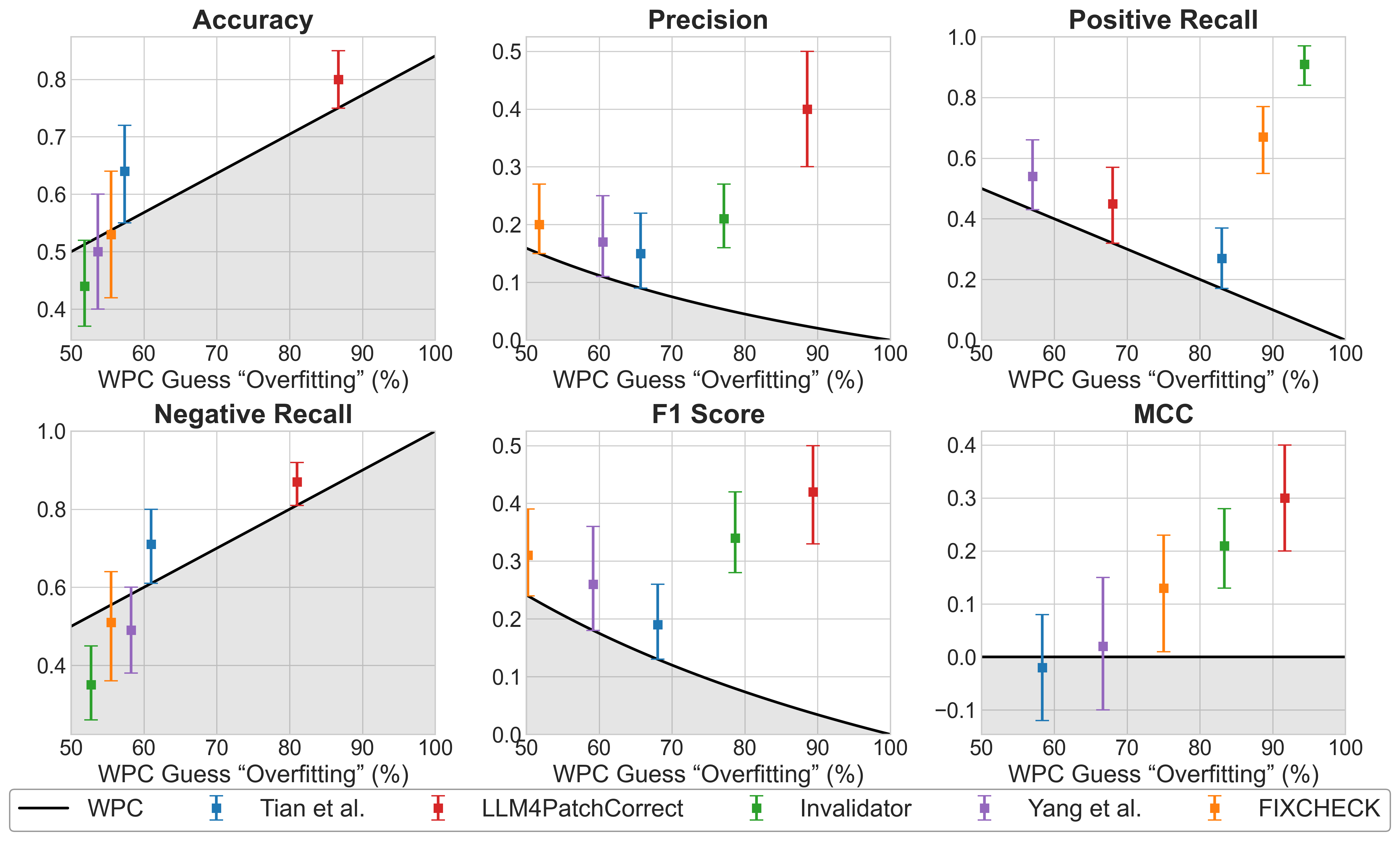}
    \caption{RQ3.
    Comparison of tool performance against the WPC baseline for the \dataOne{} dataset. Each tool’s point estimate and 95\% bootstrap confidence interval are overlaid on the WPC performance envelope, traced across prior probabilities $p$ from 0.5 to 1.0 for predicting overfitting. The envelope represents the best achievable performance of a naive, distribution-aware classifier. As each tool’s performance is independent of $p$, their corresponding points are distributed along the x-axis to highlight potential interactions with the envelope; otherwise, they are positioned in the remaining space to ensure visual clarity.}
    \label{fig:wbc-all-metrics}
\end{figure}

\subsubsection{WPC Baseline}
The comparison with RS has already shown that naive strategies set a surprisingly high bar to beat for developer patch inspection effort. The WPC baseline (Section~\ref{sec:wbc}) provides an alternative strategy by enhancing a guesser with one additional piece of knowledge: the empirical class prior. 
Figure \ref{fig:wbc-all-metrics} delineates the best performance any naive detector can achieve without looking at the patch at all, merely by tuning the prior $p \in [0.5,1.0] $ with which it predicts ``overfitting”. This distribution-aware oracle converts every quality metric into an envelope, represented by the shaded grey region. Each POD tool's performance on the \dataOne{} dataset is plotted as a point estimate along with 95\% bootstrap confidence intervals, with the metric value on the y-axis and the oracle baseline's $p$ value on the x-axis. A tool is deemed strictly superior if its lower confidence bound exceeds the envelope’s maximum across all $p$, partially superior if the lower bound intersects the envelope, and subperformant if its upper bound remains below the envelope’s minimum. We acknowledge that this baseline reaches theoretical maxima for certain metrics that no realistic tool can match (e.g., perfect negative recall at $p = 1$); yet, it still allows for the detection of robust gains.

Across the assessed metrics for the \dataOne{} dataset, Invalidator is the only technique whose lower confidence bound does not dip inside the WPC envelope on four occasions (precision, positive recall, F1, and MCC). Despite this, it is the only tool to perform worse on a metric: negative recall. In particular, its upper confidence interval of 0.450 is below the minimum value for WPC at $p = 50\%$. LLM4PatchCorrect exhibits a slightly weaker pattern: it is strictly superior on three metrics and intersects the envelope for accuracy and the recall metrics. Tian et al., Yang et al., and FIXCHECK show progressively less robust gains, with the latter two falling below the envelope in multiple metrics. 
Since a heavily skewed prior can push the WPC accuracy envelope up to 0.84, none of the tools clear the baseline for every $p$. LLM4PatchCorrect surpasses the envelope once $p < 86.7\%$, and Tian et al. does so only for $p < 58.8\%$. The remaining three tools are indistinguishable from the naive guesser in accuracy across most of the prior range, reflecting their difficulty in compensating for the majority-overfitting class.\looseness=-1

Detecting correct patches with high precision is where LLM4PatchCorrect outperforms all other approaches. 
Along with Invalidator, these tools clear all baseline values of $p$ (including confidence intervals). The other tools become reliably superior only once the prior exceeds a certain threshold.\looseness=-1

It is no surprise that the two dynamic approaches, which demonstrated superior performance in detecting correct patches, outperformed a naive guesser which favours the overfitting class in positive recall. What is interesting is that the learning-based tool of Tian et al. only outperforms a naive guesser in the retention of correct patches when the guesser's probability of classifying a given patch as correct is $< 17\%$.

The above picture flips for the majority class: a simple ``always overfitting’’ guesser achieves perfect negative recall at $p = 1$. Consequently, only the learning-based approaches which demonstrated superior performance in the correct classification of overfitting patches can improve over some portions of the baseline. As mentioned previously, Invalidator performs worse than the baseline in negative recall, even at $p = 0.5$, which effectively demonstrates the effect of a `coin flip' guesser.
F1-Score performance mirrors the same trends as for precision. Invalidator and LLM4PatchCorrect again cleared the envelope throughout, achieving average values of 0.34 and 0.42. 
As the WPC expectation for MCC collapses to zero correlation for all $p$, it is the most discriminative criterion. LLM4PatchCorrect, Invalidator, and FIXCHECK exhibit positive, significant MCCs (0.30, 0.21, 0.13, respectively), whilst Tian et al. and Yang et al. straddle zero.

The equivalent figure for the \dataTwo{} dataset can be found in our replication package~\repo. For this dataset, we observe that Invalidator is not as strong as for the \dataOne{} dataset, as it only beats the baseline for all values of $p$ on two metrics (F1 and positive recall). Similarly, LLM4PatchCorrect no longer outperforms the baseline for all values of $p$ on any metric. In fact, no tools surpassed the baseline in MCC.
Accuracy also suffers across the board, and LLM4PatchCorrect's lower confidence interval only surpasses the WPC baseline at values of $p < 61.8\%$, while all other tools fall below the baseline. Additionally, the Yang et al. tool joins Invalidator and FIXCHECK in surpassing the WPC baseline in positive recall for all $p$. However, negative recall also suffers for these tools, as all three perform worse than the baseline, indicating a tendency to consistently over-predict correctness.

Taken together, the RS (effort) and WPC (prior-aware quality) baselines expose a common gap: on both datasets, no detector simultaneously (1) keeps the mean and median inspection effort below that of random selection with 95\% confidence of finding a correct patch (RS-95) and (2) offers partial superiority over the WPC envelope across the core metrics. On the \dataOne{} dataset, Invalidator wins on inspection effort but fails to outperform WPC on negative recall, even for $p = 0.5$, whilst requiring an average runtime of more than 3 minutes per patch on our machine. LLM4PatchCorrect delivers faster performance, outperforming the WPC at least partially on all metrics, yet requires developers to inspect more patches than RS-95. The remaining tools fail to meet both requirements.

\section{Threats to Validity}
\label{Treats}
Our experiments focused on Defects4J Java projects, a set of patches generated by 11 different APR tools, and six overfitting POD techniques, each representative of a given class. Thus, it is possible that our results may not generalise to different programming languages, patches and patch overfitting detection tools. 
However, we focused on Java for this study because the majority of APR/POD tools and labelled datasets target this language~\cite{fei-2025-pca} and identified POD tools that are representative of the current state-of-the-art, recent, and open-source.
Furthermore, we mitigate possible threats due to differences in tool implementation and configuration by using their publicly available implementations or re-implementations closely guided by the original work, and we set parameters as recommended by the authors.

Regarding the datasets, we used the only two that, to the best of our knowledge, accurately represent the distributions of correct-to-overfitting patches generated by APR tools under consistent conditions (as described in Sections~\ref{sec:current-overfitting-detection-evaluation-techniques}~\&~\ref{sec:dataset}).

When using any LLM-based technique (such as LLM4PatchCorrect), data leakage is a possible threat~\cite{williams2026reflection}. 
Defects4J bugs may have been present in the underlying model's (Starcoder-7b) training data. 
However, we used labeled patch datasets that were created well after the model's training cut-off date to mitigate the threat of data leakage. 

\section{Conclusions}
We have provided a comprehensive evaluation of \totDetectors{} state-of-the-art POD approaches --- each representative of the three different classes existing in the literature --- in a \textit{practical} scenario, to provide guidance for the \textit{benchmarking} of such techniques in future work.  
 
Comparing these tools against two naive baselines, RS and WPC, our results are striking: no single technique achieves strong accuracy in detecting both classes of patches (correct and overfitting), and, in fact, most tools struggle to substantially outperform a naive random classifier.

Consequently, our findings suggest that further work is needed to devise more effective POD techniques. For example, we observed that dynamic approaches excel at identifying correct patches (while over-predicting correctness for overfitting patches), whereas learning-based methods are better at identifying overfitting patches (at the cost of discarding many correct ones). Thus, it is possible that a combination of learning-based and dynamic-based methods could offer the best of both worlds.

Based on these findings, we encourage the APR community to benchmark any novel POD tool against RS and WPC.
We encourage future work to follow the methodology presented herein, and, to further facilitate its adoption, we make all our scripts and data publicly available \repo{}.

\bibliographystyle{IEEEtran}
\bibliography{references}
\end{document}